\documentclass[10pt, conference]{IEEEtran}
\IEEEoverridecommandlockouts
\usepackage{cite}
\usepackage{amsmath,amssymb,amsfonts}
\usepackage{algorithmic}
\usepackage{graphicx}
\usepackage{textcomp}
\usepackage{xcolor}
\usepackage{hyperref}
\def\BibTeX{{\rm B\kern-.05em{\sc i\kern-.025em b}\kern-.08em
    T\kern-.1667em\lower.7ex\hbox{E}\kern-.125emX}}
\input{macro}

\begin{document}

\title{Data Driven Testing of Cyber Physical Systems \\
}

\author{\IEEEauthorblockN{Dmytro Humeniuk}
\IEEEauthorblockA{\textit{Polytechnique Montréal} \\
Montreal, Canada \\
dmytro.humeniuk@polymtl.ca
}
\and
\IEEEauthorblockN{Giuliano Antoniol}
\IEEEauthorblockA{\textit{Polytechnique Montréal} \\
Montreal, Canada \\
giuliano.antoniol@polymtl.ca}
\and
\IEEEauthorblockN{Foutse Khomh}
\IEEEauthorblockA{\textit{Polytechnique Montréal} \\
Montreal, Canada \\
foutse.khomh@polymtl.ca}
}
\maketitle

\begin{abstract}
Consumer grade cyber-physical systems (CPS) are becoming an integral part of our life, automatizing and  simplifying everyday
tasks.
Indeed, due to complex interactions
between hardware, networking and software, developing and testing such
systems is known to be a challenging task. 
Various quality assurance and
testing strategies have been proposed.

The most common approach for pre-deployment testing is to model the system and run
simulations with models or software in the loop. In practice, most often, tests are
run for a small number of simulations, which
are selected based on the engineers’ domain knowledge and experience.



In this paper we propose an approach to automatically generate fault-revealing test cases for CPS. We have implemented our approach  in  Python, using standard frameworks and used it to generate scenarios
violating temperature constraints for a smart thermostat implemented as a part of our IoT testbed. 
Data collected from an application managing a smart building have been used to learn  models of the environment
under ever changing conditions. The suggested approach 
allowed us to identify several pit-fails, scenarios (\ie  environment conditions and inputs),
where
the system behaves not as
expected. 





\end{abstract}


\section{Introduction{\label{sec:introduction}}}
Ensuring the reliability of cyber-physical systems (CPS), such as cars with adaptive cruise control
, robotic systems or smart buildings is of vital importance. Testing software for such systems is challenging, as developers need to take into account the interaction with hardware as well as the environment.
Consider an autonomous vehicle that may drive in various conditions (fog, rain, snow) and react to pedestrians as well as other cars manoeuvres. Or an IoT network, where, due to extensive traffic, control commands would arrive with delay or get lost.

The input search space of such systems is substantial, therefore metaheuristics and random search based techniques, are often used to generate the test cases \cite{b1}.
Further, the system model is used to execute the test cases, as it is unpractical to use the physical system, especially on the pre-deployment testing stage \cite{b2}. Therefore, it is also important to obtain an accurate system model, that is not time consuming to execute.

A number of tools have been developed to verify whether a system, represented by its model, meets specific requirements and whether there are any inputs  violating them, e.g., S-Taliro, Breach, falsify ARIsTEO and other tools described in \cite{ernst2020arch}.
However, they require to manually specify the requirements, which is a tedious task and does not guarantee the consideration of all possible requirements.

Another direction of research is automatic search based generation of test suites for CPS.
Those approaches are mostly often focused on finding the test cases with the best requirements coverage and diversity, but not falsification \cite{arrieta2017search, matinnejad2016automated,b1}. They lack flexibility as often require an external software to generate the initial test cases. Also, they are using the Simulink API to execute the models, which can be computationally expensive.

We surmise that it is important to design test suites with high fault revealing power, indicating to developers the possible worst case scenarios of system execution. Moreover, the test cases should consider possible combination of environmental conditions during system execution. For example, a car can ride on a dry or an icy road, changing the model describing it's trajectory evolution.

\textbf{Motivation}. We explain the motivation for our work with a wirelessly controlled thermostat case study. This system is described in a greater detail in \cite{zid2020double}. The thermostat automatically controls the temperature in a closed room by switching between "on" and "off" modes. The temperature in the room is set by the user defined schedule. The developer writing software for the thermostat defines a number of parameters such as the sampling rate, the hysteresis value (small threshold before or after reaching the temperature), etc. In addition, the environmental conditions affect the system behaviour: the commands sent from the controller to the thermostat can be delayed or even lost due to the network overload, the temperature decrease/increase speed can vary depending on the time of the day, room humidity, etc. Considering all the parameters, are there scenarios when the system is not able to follow the schedule? What are the combinations of input parameters and environmental conditions that drive the system to an unsafe state?
 Finding answers to such questions motivated us in developing an automatic search based approach for CPS test case generation.
 


 \section{Problem formulation{\label{sec:problem}}}

The behaviour of a hybrid system can be described with modes having continuous output dynamics and discrete mode switches \cite{alur2015principles} . Each hybrid system has input(s) $U_i$, output(s) $Y_i$ and state variables $S_i$. The expected system behaviour $B(\tau)$ is specified over a time interval $\tau$. 
Mode switch occurs when the expected output requirements can’t be met by the system in a particular mode. For example, when the temperature is set to be lower, than the current room temperature, the thermostat will switch from heating to cooling mode. The dynamics of system state, input and output variables $S_i$, $U_i$, $Y_i$ in each mode $N_i$ is given by a corresponding mathematical model $M_i$, which can be derived from system execution data using system identification or machine learning techniques \cite{menghi2020approximation}.  
Being dependent on the environmental conditions, the models can get very complex and taking a substantial amount of time to execute. We therefore surmise that each mode $N_i$ can be represented by a family of surrogate or simplified models $M_i$, corresponding to certain environmental conditions $E_i$. Therefore each family $M_i$ will contain models $m_{ij}$ , where $i$ is the model family identifier and $j$ - model identifier.
The test case (TC) generation problem can thus be thought of as finding  a combination of models $m_{ij}$ and input values $U_i$ maximizing the difference between simulated system behaviour $B_s(TC)$ over time interval $\tau$ and expected behaviour $B(TC)$, with system variables satisfying a certain constraint $K$:
\begin{gather*}
    \max{\delta(B(TC), B_s(TC))}, \\
    subject \;to:
K(Si, Ui, Yi).
\end{gather*}

Where $\delta(B(TC), B_s(TC))$ is our fitness function computing the deviation between the expected and simulated behaviour. 
The expected behaviour should be defined in the test cases, such as, for instance, a temperature schedule or a car trajectory to follow. Typically, these values will be used as the input to the model. Therefore, the fitness function calculates how a system output, as a reaction to a particular input, is different from the expected output. 
The constraint can be represented by an acceptable range of input values, duration of the simulation, etc. By using different families of models of the system, we account for its behaviour in different environmental conditions.
Evolution can be done in one way: for the fixed scenarios find a combination of models, violating the user requirements. Or in both ways: by changing models and system inputs, find the worst possible scenario.

To generate the initial test cases we suggest using hidden Markov chains.It can be created automatically, given the system modes and probabilities of switching between them. This idea is not new, in \cite{liu2019modeling} for example, Markov chains are used to generate simulation scenarios for a wireless network. 
We decide to use the Markov chain for two main reasons. First, by running the chain for a number of times, the developer can estimate an average performance of the system. Secondly, in our experiments, the initial population for genetic algorithm (GA) generated with Markov chain provided semantically better test cases, than completely random initialization. 

The parameters for the Markov chains, such as states and probabilities of state change, can be estimated from the data on typical system usage scenarios. In this case, "states" correspond to the system modes. For each state, there is a set of possible output values the system can reach, the duration of being in the state and the model, accounting for the system behaviour corresponding to particular environmental conditions.
Therefore, for scenario generation, operation in each state can be represented by a triplet:
\begin{equation}
 S_i = (Y_i, \tau_i, M_i)
\end{equation}
where $Y_i$ is desired system output in a particular state, $\tau_i$ - the duration of this state (the sum of $\tau_i$ should be equal to $\tau$) and $M_i$ - model family to use, to describe the system behaviour. The duration can be specified in time units, or other units, such as total length, as in case of generating road trajectories.
A test case is represented by a sequence of states:
\begin{equation}
 TC_i = (S_1, S_2, ..., S_i)
\end{equation}
 Finally, we suggest using evolutionary algorithms to find the combination of the $Y$, $\tau_i$ and $M_i$ values maximizing the fitness function $\delta$.
 For this study we used a single objective genetic algorithm.

 \section{Case study{\label{sec:case}}}
In our case study we consider an example of a wirelessly controlled thermostat, described in the Introduction.

For the thermostat two modes of operation can be defined: $M_1$: “ON” and $M_2$: “OFF”. The behaviour of the system can be represented by a sequence of switching between these modes, and time  passed in each mode.
The input variable $U$ is the goal temperature (expected behaviour) at a given point in time. The output variable $Y$ is the value of output temperature controlled by the system. It can also have such state variables as $S_1$ - system start temperature in  a certain mode, $S_2$ - time spent in a particular mode, etc.
The constraints for the input variables  are the temperature values between 16 and 25 degrees Celsius. The time intervals spent in each mode can range from 15 minutes to 6 hours.

\subsection{Model creation}
To create the system model we used a system identification technique \cite{ljung1994modeling}, where a model of a dynamical system is built from the data. The process requires the following steps:
\begin{enumerate}
    \item Extract the data describing system behaviour in different modes.
    \item Select a model structure.
    \item Apply an estimation method to estimate values for the adjustable coefficients in the candidate model structure.
    \item Evaluate the estimated model.
\end{enumerate}
The wireless thermostat, controlling  temperature in a closed room, is a part of our physical testbed of IoT network of more than 30 devices, based on a Z-wave protocol \cite{zid2020double}. Therefore, we extracted the data for creating the model from the experimental measurements. We selected the series of data points, corresponding to behaviour of the thermostat after "switch on" and "switch off" commands.
One model includes two equations describing behaviour in "on" and "off" modes. In total, we could identify 15 models having different coefficients in the equations. Evidently, due to varying environmental conditions, i.e the opened door, higher or lower humidity, heat transfer from outside, the coefficients in the selected model structure had to be adjusted to better fit the original data. Creating one complex model, with high number of inputs, considering the environmental conditions, would make the execution of the model computationally expensive.

One of the challenges is to select the model structure. In our case, the heating and cooling of a closed space is guided by physical laws, such as Newton Law of cooling \cite{winterton1999newton}. The law has an exponential nature, therefore our experimentally selected model structure is based on increasing and decreasing exponential function.

We propose the following time-discreet model structure for the $M_1$ ("on") mode:
\begin{equation}
    Y = k_{on1}*(1 - e^{-k_{on2}*t_i}) + T_0
\end{equation}
and for the $M_2$ ("off") mode:
\begin{equation}
    Y = k_{off1}*(e^{-k_{off2}*t_i}) + T_0 - k_{off1}
\end{equation}
Here $k_{on1}$, $k_{on2}$, $k_{off1}$, $k_{off2}$ are the unique coefficients defining the model behaviour in a particular environment. $T_0$ - is the starting temperature and $t_i$ - the discreet time step value. 
We keep the coefficients in a table, such as table \ref{tab:tab1}. As an example, we show coefficients for the three obtained models. 
\begin{table}[ht]
\begin{center}
\caption{Model coefficients}
\label{tab:tab1}
\scalebox{1}{
\begin{tabular}{|c|c|c| c | c|} 
\hline 
Model & $k_{on1}$ & $k_{on2}$ & $k_{off1}$ & $k_{off2}$\\
\hline
1 & 6 & 0.14170703 & 4.3 & 0.09531917\\
\hline
2 & 7.9 & 0.11180434 & 5.2 & 0.04803319\\
\hline
3 & 7 & 0.13425024 & 3.8 & 0.07661568\\
\hline
\end{tabular}}
\end{center}
\end{table}

To obtain the coefficients, the points from the data must be fitted by a curve with minimal deviation. We used python SciPy library, namely $curve\_fit$ function from $Optimize$ class, which  is based on non-linear least squares method \cite{scipy}.
The average root mean square error between original and approximated data did not exceed 0.5 degrees.
\subsection{Generating initial test cases}
\begin{figure}[ht]
\includegraphics[scale=0.22]{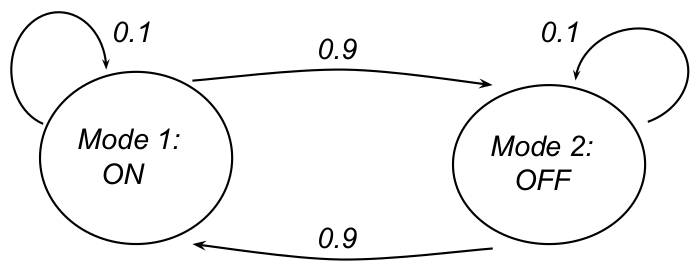}
\centering
\caption{Thermostat system representation with a Markov chain}
\centering
\label{fig:markov}
\end{figure}

To automatically generate the test cases we represent the thermostat system as a Markov chain with two states "on" and "off", which is shown in Fig.\ref{fig:markov}. The probabilities of changing the states were estimated empirically, so that most of the generated test cases are semantically correct. A change of state occurs with probability of 0.9 and state remains the same with 0.1 probability.

After reaching a particular state, we randomly choose a temperature value the system is expected to reach, the time interval to be spent in the state and the the model coefficients to use, so that each state is represented by a triplet (temperature, duration, model), similar to (1). In this way, a test case represents a temperature schedule a user might define.

For each execution we indicate the expected duration of the test case as well as the number of states. We chose the duration to be 24 hours, representing one day, and having 5 to 12 states in each test case.

We implemented the algorithm in a python script, which saves the generated test cases in a "json" format.
\subsection{Genetic algorithm description}
To find the test cases maximizing the difference between the expected and simulated behaviour we implemented a genetic algorithm in Python with Pymoo framework \cite{pymoo}. 
In our configuration the number of generations is 
$N_G$ = 90,  mutation rate is $m_r$ = 0.4, crossover rate is $c_r$ = 0.9 and
population size: $p_{size}$ = 100.
These values were established experimentally and following the common practices.
\subsubsection{Solution representation}
\begin{figure}[ht]
\includegraphics[scale=0.29]{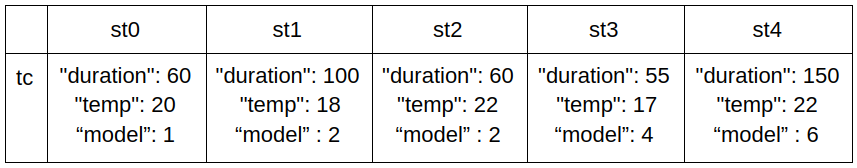}
\centering
\caption{GA chromosome representation, tc - test case, st - state}
\centering
\label{fig:chrom}
\end{figure}

The solution is composed by at least one test case, containing from 5 to 12 states.
The chromosomes are the test cases, represented in the software implementation as a dictionary, see Fig.\ref{fig:chrom}. They have a variable number of genes, where each gene corresponds to a system state. 
In the figure, "temp" parameter describes the expected temperature in each state, corresponding to the system input values, "model" - a family of models to use and "duration" - time the system should spend in each state.
\subsubsection{Selection}
We use the k-way tournament selection implemented in Pymoo to choose the parents.
\subsubsection{Crossover operators}
\begin{figure}[ht]
\includegraphics[scale=0.25]{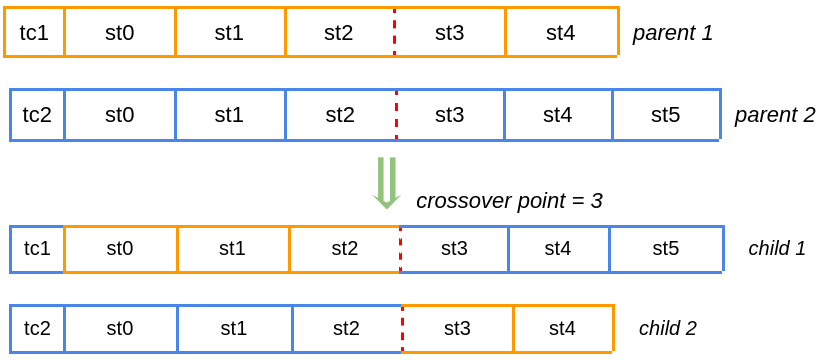}
\centering
\caption{Crossover operator for two test cases of 5 and 6 states with crossover point at the third state}
\centering
\label{fig:cross}
\end{figure}

We implemented a crossover operator that exchanges the states between two different test cases as shown in Fig.\ref{fig:cross} We use a one point crossover.
\subsubsection{Mutation operators}
We define two mutation operators, similar to \cite{arrieta2017search}:
\begin{itemize}
    \item \textit{exchange operator}: two states of a chromosome are randomly selected and exchanged the positions;
    \item \textit{change of variable operator}: a state in a chromosome is randomly selected, then for one of the state variables (temperature, duration, model) value is changed according to its type and maximum as well as minimum values.
\end{itemize}

\subsubsection{Fitness function}
In our study the fitness function evaluates the root mean square error between the simulated and expected behaviour. The expected behaviour is specified in the test cases, which are given as an input to the system simulation. The test case is executed using the specified models and the values of system behaviour are calculated. The fitness of 1 signifies that the system can provide the temperature with the difference from the schedule of 1 degree on average, which might be acceptable for a typical user. As the Pymoo framework minimizes the fitness function, in our implementation we multiply its actual value by (-1).

 \section{Results{\label{sec:results}}}

\begin{figure}[ht]
\includegraphics[scale=0.25]{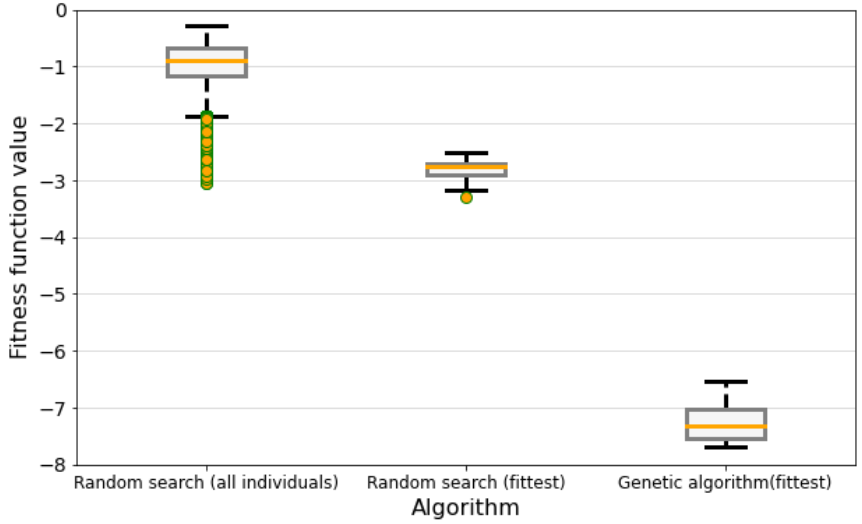}
\centering
\caption{Genetic algorithm and random search comparison after 50 runs}
\centering
\label{fig:box}
\end{figure}

\begin{figure}[ht]
\includegraphics[scale=0.23]{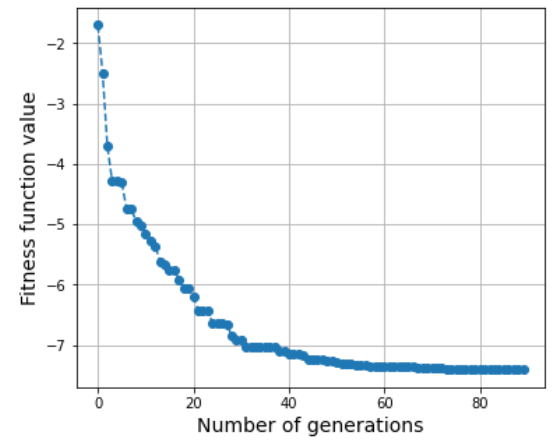}
\centering
\caption{Convergence of GA over 90 generations}
\centering
\label{fig:conv}
\end{figure}

To evaluate the performance of our GA implementation we ran it 50 times (each run contains 9000 evaluations). After each run we recorded the fittest individuals. We compared its performance with the random search (RS).
We recorded the fittest individual after generating 9000 random individuals, repeating the process 50 times. The obtained boxplot is shown in the Fig. \ref{fig:box}. In the boxplot we also report the fitness values of all randomly generated individuals during evaluation. We can see that GA always produces better results with an average fitness of -7.2, while the average fitness of the RS best individuals is -2.8. Considering all the randomly generated individuals, the average fitness is around 0.93.
For one of the runs we also report the convergence of GA in Fig. \ref{fig:conv}, which confirms its good performance.

From this evaluation we conclude that our thermostat system performs well on average (the mean deviation from the schedule is around 1 degree) as shown by all randomly generated schedules. However, there are potential scenarios, which can lead to completely wrong system behaviour (deviation from the schedule for 7 degrees on average). It is up to developer to decide, whether the found test cases are realistic or not. If they aren't, we recommend adjusting the search parameters and constraints.

 \section{Discussion and conclusion{\label{sec:conc}}}
In this paper we suggested an approach for generating fault revealing test cases for hybrid CPS, taking into account variability of system behaviour in changing environmental conditions. It includes generation of models, initial test cases and genetic algorithm implementation in Pymoo framework. The results for the wireless thermostat case study prove the effectiveness of our implementation comparing to  random search. With our approach we could evaluate the system performance as well as generate potentially dangerous scenarios.
However, it is up to developers to judge if the test cases are pertinent and take further actions to prevent the failures.

The approach can be applied for a wide range of hybrid CPS, what we are going to demonstrate in our future case studies.
We also plan to implement our approach as a complete test case generation tool.


\bibliographystyle{unsrt}
\bibliography{refs} 

\end{document}